\documentclass[preprint2]{aastex}

\shorttitle{Neutron-rich matter in bubbles in SNe}
\shortauthors{Hoffman et al.}
\begin{document}
\bibliographystyle{apj}
\title{Nucleosynthesis in early supernova winds III: \\No significant contribution from neutron-rich pockets} 
\author{R. D. Hoffman, J. Pruet \& J. L. Fisker}
\affil{Lawrence Livermore National Laboratory, P.O. Box 808, Livermore, CA  94550}
\email{hoffman21@llnl.gov,pruet1@llnl.gov,fisker1@llnl.gov}
\author{H.-T. Janka \& R. Burras}
\affil{Max-Planck-Institut f\"ur Astrophysik, Karl-Schwarzschild-Strasse 1, 85741 Garching, Germany}
\author{S. E. Woosley}
\affil{Department of Astronomy and Astrophysics, University of California, Santa Cruz, CA 95064}
\begin{abstract}
Recent nucleosynthesis calculations of Type II supernovae using advanced neutrino
transport determine that the early neutrino winds are proton-rich.
However, a fraction of the 
ejecta emitted at the same time is composed of neutron-rich pockets. 
In this paper we calculate the nucleosynthesis 
contribution from the neutron-rich pockets in the hot convective bubbles of 
a core-collapse supernova and show that they do not contribute significantly 
to the total nucleosynthesis.
\end{abstract}


\section{Introduction}\label{sec:introduction}
During a delayed Type II supernova explosion, the collapsing core emits neutrinos 
and anti-neutrinos. These cool the shrinking proto-neutron star and heat the 
infalling matter which expands outwards, reverses the in-going accretion shock, 
and hypothetically causes the supernova to explode. The heating is sufficiently 
rapid to establish and maintain a convective region between the infalling matter 
and the proto-neutron star. The matter --- originally part of the progenitor's 
silicon burning shell --- in this convective region comprise electrons, positrons, 
and completely photo-disintegrated nuclei (protons with a mass fraction, $Y_e$, 
neutrons with mass fraction $1-Y_e$). The neutrinos irradiate the convectively 
overturning bubbles, so this matter is not simply ``adiabatically expanding'' nor 
is it subject to a uniform history of neutrino irradiation. This is important because 
$\nu_e+n \rightleftarrows e^-+p$ and $\overline{\nu}_e+p \rightleftarrows e^++n$ 
reactions along with artificially boosted \citep{Janka03} neutrino-luminosities drive 
the matter proton-rich due to the lighter proton mass \citep{Pruet05,Froehlich06b,Froehlich06a} 
given approximately equal neutrino luminosities of the neutrinos and anti-neutrinos 
\citep{Liebendoerfer03}.
Thus different pockets in the bubbles will have different compositions and different 
$Y_e$, some of which are neutron-rich.

The contribution to the nucleosynthesis of the proton-rich bubbles was investigated 
by \cite{Pruet05} and the contribution to the nucleosynthesis of the proton-rich winds 
was investigated by \cite{Pruet06}. Both calculations were based on the Lagrangian 
$(\rho,T)$-histories of tracer particles in the 2D model of \cite{Janka03}. 
However, some bubbles also 
contained neutron-rich pockets which whose nucleosynthesis was not explored in those papers. 
This is the 
subject of this paper.

We have extracted tracer particle trajectories for these neutron-rich pockets 
and investigate their 
nucleosynthesis contribution to the overall ejecta. 
In the following, \S \ref{sec:model} describes the 
supernova model and the $Y_e$-distribution of matter in more detail. The 
nucleosynthesis results are given in \S \ref{sec:synthesis}  followed by a 
conclusion in \S \ref{sec:conclusion}.

\section{Supernova model}\label{sec:model}
Our calculations of the nucleosynthesis contribution of neutron-rich pockets use the 
same supernova model as \cite{Pruet05,Pruet06} but here we consider the 
$(\rho,T)$-trajectories with $Y_e<0.5$ 
 thus complementing our earlier calculations.

The model is described in of \cite{Janka03} (see \cite{Rampp02} for specific code 
details and \cite{Pruet05} for more details). In this model, the progenitor is based 
on a non-rotating $15M_\odot$ model (S15A) of \cite{Woosley95} which is transferred 
to a 2D polar grid (400 non-equidistant radial zones and 32 poloidal zones) using 
random velocity perturbations of the order of $10^{-3}$ to seed the convection and an 
artificial 20--30\% enhancement of the neutrino flux to ensure the supernova explosion. 

The simulation commences at $t_i=-175\,\textrm{ms}$ prior to the core bounce and 
uses embedded tracer particles to provide a history of $(\rho,T,Y_e)$ for a range of 
electron abundances until $t_f=470\,\textrm{ms}$ after the core bounce at which time 
the 2D simulation was stopped due to CPU-constraints. 

At $t_f$, the temperature is still several billion K so the nucleosynthesis is still in 
partial statistical equilibrium and not yet frozen out. To continue the nucleosynthesis 
calculation, the density and the temperature was mapped from the 2D model to a 1D grid 
and extrapolated by assuming a homologous expansion 
with a constant electron abundance and a constant entropy. These assumptions are 
acceptable for calculating the $(T,\rho)$-response to the subsequent expansion 
since the nuclear decays are too slow to change $Y_e$ over the expansion timescale. 
Also the rate of expansion is so large that the ``$r^{-2}$''-dependent neutrino-luminosity 
quickly becomes irrelevant \citep{Pruet05}.

\begin{figure}[tbph]\center\includegraphics[width=1.0\linewidth,clip]{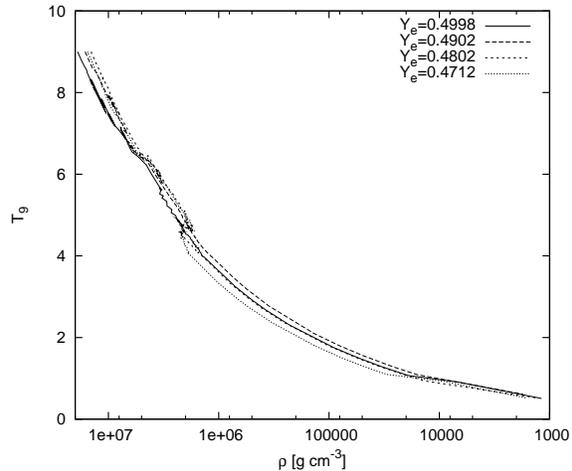}
\caption{This figure shows $\rho$ vs $T_9(\equiv10^9\,\textrm{K})$ for some representative $Y_e$ trajectories. The transition to the extrapolation from the 2D simulation happens around $T_9=$4--5.}\label{fig:rhot}
\end{figure}

Fig.~\ref{fig:rhot} shows 4 representative trajectories of $\rho$ and 
$T_9(\equiv10^9\,\textrm{K})$ out of the 40 neutron-rich trajectories that were 
tracked during the simulation and subsequently extrapolated to lower temperatures.
The transition to the extrapolation from the 2D simulation happens around
$T_9=$4--5. The entropy is approximately $15 k_B/\textrm{nucleon}$.

\section{Neutron-rich nucleosynthesis}\label{sec:synthesis} 
In this supernova model, the amount of matter with $Y_e<0.47$ is 
$\lesssim 10^{-4}M_\odot$. This prevents an unacceptable overproduction of $N=50$ 
nuclei \citep{Hoffman96}. 

In the following we consider the nucleosynthesis in the $0.47<Y_e<0.50$ range 
($M=5\times10^{-3}M_\odot$) using the trajectories described above.
Nucleosynthesis calculations commence at $T_9=9.0$ and proceed until freeze-out 
below $T_9\sim 1$. At $T_9=9.0$, the matter comprises protons with a mass fraction, 
$Y_e$, and neutrons with mass fraction $1-Y_e$. 
Therefore the initial conditions is completely defined by the initial values of 
$\rho$, $T$, and $Y_e$. Between $T_9\sim9$ and $T_9\sim6$, ${}^4\textrm{He}$ 
quickly recombines which depletes the neutrons and protons equally thus keeping 
$Y_e$ constant. At $T_9\sim$ 4--6, the helium recombines into the iron group 
elements along the $N=28$ isotone and then forms $Z=28$ isotopes as the temperature 
drops to $T_9\sim$ 2--3. The electron-abundance or neutron to proton ratio determines 
the subsequent reaction flow. 

For $Y_e$ closer to 0.5, primarily ${}^{56,57,58}\textrm{Ni}$ are formed. 
The flow from these nuclei leads to ${}^{64}\textrm{Ge}$. Unlike the $\nu p$-process 
\citep{Froehlich06a}, there is not a sufficient amount of protons left at this 
time for neutrinos to provide a sufficient number of neutrons to capture on 
${}^{64}\textrm{Ge}$ and thus move beyond this waiting point. As a result, 
heavier isotopes are not co-produced with the ${}^{62}\textrm{Ni}$ and 
${}^{64}\textrm{Zn}$. In particular, there is no production of 
the light $p$-nuclei for $Y_e\sim 0.5$.

For $Y_e$ closer to 0.47, primarily ${}^{58,59,60}\textrm{Ni}$ are formed. 
This means that the ${}^{64}\textrm{Ge}$ waiting point is easily circumvented 
which leads to overproduction of ${}^{74}\textrm{Se}$, ${}^{78}\textrm{Kr}$, 
and ${}^{92}\textrm{Mo}$ which is co-produced with ${}^{64}\textrm{Zn}$. 
With increasing $Y_e$, the ${}^{92}\textrm{Mo}$ production falls off \citep{Hoffman96}.

\subsection{Production factors}
The total nucleosynthesis contribution is given by the sum of the mass weighted production factors
$P(i)$, defined as 
\begin{equation}
P(i)=\sum_j\frac{M_j}{M_\textrm{e}}\frac{X^i_j}{X_\odot^i}\,,
\end{equation}
where $M_j$ is the mass in the $j$th bin (trajectory), $M_\textrm{e}=13.5M_\odot$ 
is the total mass ejected in the supernova explosion, $X^i_j$ is the mass fraction of 
the $i$th isotope in the $j$th bin and $X_\odot^i$ is the solar abundance of the 
$i$th isotope taken from \cite{Lodders03}.

\begin{figure*}[tbph]\center\includegraphics[width=0.93\linewidth]{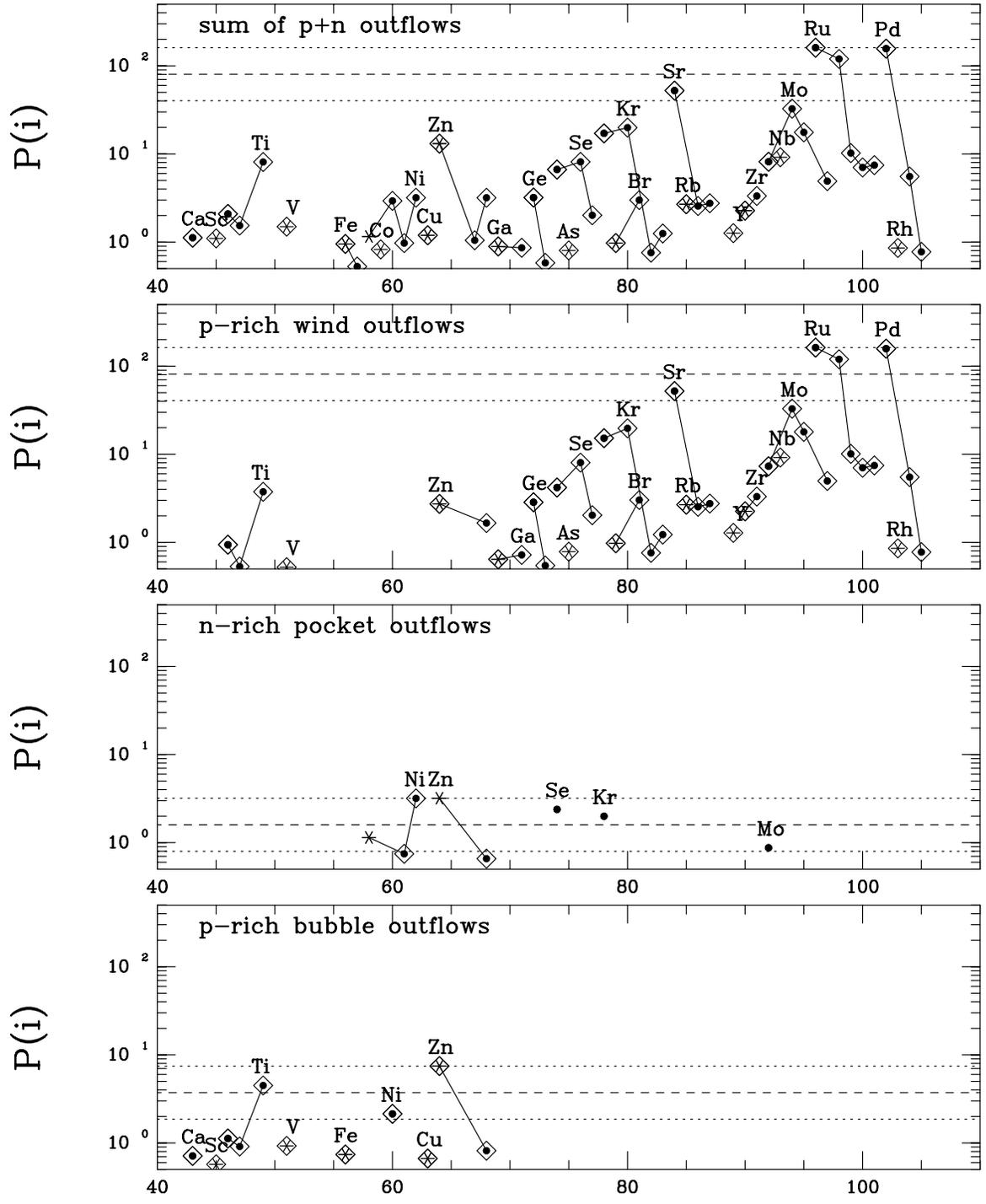}
\caption{Production factors of the neutron-rich trajectories of the convective 
bubble ejecta. The most abundant isotope for a given element is shown with an 
asterisk. Diamonds indicate that the isotope was made primarily as a radioactive 
progenitor.}\label{fig:nrbubble}
\end{figure*}

The production factors for neutron-rich pocket trajectories are shown in 
Fig.~\ref{fig:nrbubble}. The most produced isotopes in the neutron-rich parts 
of the bubble relative to solar abundances are ${}^{62}\textrm{Ni}$ and 
${}^{64}\textrm{Zn}$ which originate in pockets with $Y_e$ closer to 0.5. 
These are co-produced along with ${}^{74}\textrm{Se}$ and ${}^{78}\textrm{Kr}$ 
which originate in the pockets with $Y_e$ closer to 0.47.

The figure also shows the contributions from the proton-rich bubble 
and the proton-rich winds trajectories (emitted later). We note that the contribution 
of the neutron-rich pocket outflow is insignificant compared to the total outflow. 
The neutron-rich pockets add ${}^{74}\textrm{Se}$, ${}^{78}\textrm{Kr}$, and 
${}^{92}\textrm{Mo}$ to the bubble-outflow, but this contribution is much smaller 
than the contribution from the proton-rich winds when neutrino interactions are included. 
The neutron-rich pockets also add ${}^{62}\textrm{Ni}$ and ${}^{64}\textrm{Zn}$ 
to the total outflow but only in comparable amounts to the wind and
proton-rich pockets outflows. Here ${}^{64}\textrm{Zn}$ production is increased 
by $\sim30\%$ while ${}^{62}\textrm{Ni}$ production is increased by a factor 1.5.

\section{Conclusion}\label{sec:conclusion}
Our results show that the overproduction factors of the neutron-rich pockets 
folded with the mass-ejecta does not contribute significantly to the nucleosynthesis 
of the light $p$-nuclei of compared to the nucleosynthesis of the proton-rich pockets and winds.

\begin{acknowledgments}
This work was performed under the auspices of the U.S. Department of Energy by 
Lawrence Livermore National Laboratory in part under Contract W-7405-Eng-48 and in part under Contract DE-AC52-07NA27344. It was also supported, in part, by the DOE-OS SciDAC 
program (DC-FC02-01ER41176), the National Science Foundation (AST-02-06111), 
and NASA (NAG5-12036) and, in Germany, by the Research Center of Astroparticle Physics (SFB 375) and the Transregional Collaborative Research Center for Gravitational Wave Astronomy (SFB-Transregio 7).
\end{acknowledgments}


\end{document}